\documentclass{natureprintstyle}

\usepackage{amsmath}
\usepackage{amssymb}
\usepackage{graphicx}
\usepackage{subfigure}
\usepackage{cite}
\usepackage{color}
\usepackage{enumerate}
\usepackage[right]{lineno}
\usepackage{indentfirst}
\usepackage{fancyhdr}

\makeatletter
\renewcommand{\@biblabel}[1]{\quad#1.}
\makeatother

\graphicspath{{figures/}}

\date{}

\begin{document}

\title{Correlation between social proximity and mobility similarity}
\author{Chao Fan $^{1,2}$,
Yiding Liu $^{1}$,
Junming Huang $^{1,3,\ast}$,
Zhihai Rong $^{1,4}$,
Tao Zhou $^{1,4}$}
\maketitle

\begin{affiliations}
$^1$ CompleX Lab, Web Sciences Center, University of Electronic Science and Technology of China, Chengdu, 611731, P.R. China;\\
$^2$ College of Arts and Sciences, Shanxi Agricultural University, Taigu, 030801, P.R. China;\\
$^3$ Center for Complex Network Research, Department of Physics, Northeastern University, Boston, MA 02115, USA;\\
$^4$ Big Data Research Center, University of Electronic Science and Technology of China, Chengdu 611731, P.R. China.\\
$\ast$ E-mail: mail@junminghuang.com
\end{affiliations}


\abstract{
Human behaviors exhibit ubiquitous correlations in many aspects, such as individual and collective levels, temporal and spatial dimensions, content, social and geographical layers. With rich Internet data of online behaviors becoming available, it attracts academic interests to explore human mobility similarity from the perspective of social network proximity. Existent analysis shows a strong correlation between online social proximity and offline mobility similarity, namely, mobile records between friends are significantly more similar than between strangers, and those between friends with common neighbors are even more similar. We argue the importance of the number and diversity of common friends, with a counter intuitive finding that the number of common friends has no positive impact on mobility similarity while the diversity plays a key role, disagreeing with previous studies. Our analysis provides a novel view for better understanding the coupling between human online and offline behaviors, and will help model and predict human behaviors based on social proximity.}

\rm
\setlength{\parindent}{2em}

\section*{Introduction}
~\\


Understanding the trajectories of human daily mobility and the underlying mechanism governing their patterns is important for traffic control\cite{meyer1984urban, barthelemy2011spatial}, city planning\cite{makse1995modelling, yuan2012discovering}, public health\cite{vespignani2009predicting, wesolowski2015impact}, disaster relief\cite{Lu2012Predictability} and mobile marketing\cite{bauer2005driving, scharl2005diffusion}. Due to the lack of data, human mobility was approximately described by random walk, L\'evy fight or other diffusion processes in the past\cite{Brockmann2006Scaling,han2011Origin}. However, in recent years, the availability of massive digitized human behaviors allows quantitatively investigating human mobility and its associated factors such as living area\cite{Gonzalez2008Understanding}, personal preference\cite{pappalardo2015returners}, social contacts\cite{Cho2011Friendship}, etc. Scientists have uncovered various new statistical characteristics of human mobility unlike traditional cognitions, including heavy-tailed distribution of inter-event time and displacement distance, spatially confined living region, high regularity, periodicity and predictability\cite{Brockmann2006Scaling, Gonzalez2008Understanding, Song2010Modelling, Song2010Limits}.

The newly appeared data sources, like bank note, mobile phone, vehicle Global Positioning System (GPS) and Location-based Social Network (LBSN), make it possible to combine the study of human mobility with location context and social relationship. For example, the explosion of online social network services provides a platform for various kinds of purposes, such as making friends, getting information, or even hunting jobs. It gains increasing interests in academia and industry to explore the mutual influence between human mobility and social connections, which has significant value for location prediction, location recommendation, friend recommendation, etc\cite{lian2015mining, gao2014data}. Previous studies reveal that human offline behaviors are highly correlated with their closely connected online friends. Theoretically, a study based on mobile phone records reveals a scaling relationship between human mobility and communication pattern\cite{deville2016scaling}. Empirically, it has been found that social connections drive $10\%-30\%$ daily trips, especially for the long-distance travels\cite{Cho2011Friendship, grabowicz2014entangling}. Geographically closely lived individuals are more probable to be online friends than those live far apart\cite{liben2005geographic}. People tend to maintain social contacts when they change their living environment or migrate away\cite{saramaki2014persistence, jo2014spatial}, even after disasters like earthquake\cite{Lu2012Predictability}. Besides, researchers find that the mobility similarity between individual trajectories is correlated with social proximity, revealing the relation between moving pattern in physical space and social network structure in cyber space. Online connected individuals are more similar in mobility than disconnected ones\cite{toole2015coupling}. Symmetrically, the more similar individuals are in mobility patterns, the more frequent they maintain online communication, and the more common friends they have online\cite{Wang2011Human}. Previous researchers also verified that the micro-structure of social networks has significant impact on social contagion and relationship~\cite{ugander2012structural, ugander2013subgraph, backstrom2014romantic, weng2013virality}, suggesting that there are great differences of intimacy between friends even they have the same number of common friends. However, there are still many open questions worthy of further study, such as whether there are better indicators to measure the proximity and similarity between two individuals, whether the structure of social network impacts on one's mobility trajectory, and how to find pairs of individuals with similar behavior patterns.

In this report we perform a finer analysis to demonstrate what and how social proximity measurements are correlated with mobility similarity between two individuals based on an LBSN dataset (see Methods), in which people share real-time locations (usually referred to as "check-in") with online friends. Compared with other kinds of data sources, LBSN data have the properties of large-scale mobile records, annotated locations with descriptive tags, user-driven sparse data and explicit social friendship\cite{gao2014data}. The LBSN dataset offers a bridge connecting social network and mobility trajectories. Specifically, there are many types of online social connections can be used to measure social proximity according to different criteria, e.g., whether two individuals are friends, whether they have common friends, and how the common friends are connected. It is well studied in complex networks that "befriending" and "having common friends" are strongly related\cite{lu2011}. Besides, people visit different locations in their daily life for working, living, entertainment, etc. Two randomly selected individuals may behave similar or different visiting patterns in the physical space. It has been observed that a variety of demographic attributes, such as gender, age, education background and job are highly correlated with mobility tracks\cite{yan2013diversity, zhong2015you, lenormand2015influence}. A pair of friends with same personal attribute has a higher probability to behave more similarly. In our work, we find that human mobility similarity is strongly correlated with the existence of social connection and common neighbors (common friends on social network). Once the existence of online connection and common neighbors is given, the number of common neighbors has no positive impact on mobility similarity, while the higher diversity in common neighbors brings higher similarity in mobility pattern.

\section*{Results}
~\\

The mobility similarity between a pair of individuals is measured with \emph{Spatial Cosine Similarity} ($SCos$), which is the cosine similarity of two individuals' trajectory vectors (see Methods). Obviously, a higher $SCos$ indicates more similar behaviors. Besides, four metrics are used to measure social network proximity: (i) whether two individuals are friends; (ii) whether they have common neighbors; (iii) how many common neighbors they have; and (iv) the number of connected components in the induced subgraph by these common neighbors.

We start by investigating the effect on mobility similarity of two social network proximity metrics, namely "befriending" (whether two individuals are friends) and "having common friends" (whether they have common neighbors on social networks). Figure 1(a) and (b) report the probability distributions of mobility similarity between pairs of friends and non-friends (see Methods), and pairs of friends with or without common neighbors respectively. Figure 1(c) further reports the expected similarity of four configurations: whether or not two individuals are friends and whether or not they have common neighbors. From Fig. 1(a) we know that the pairs of friends are observed with constantly higher mobility similarity than non-friends, i.e. 0.03084 (friends) versus 0.00049 (non-friends) on average of mobility similarity. Similar phenomenon is observed in Fig. 1(b), i.e. average mobility similarity between friends with common neighbors (0.04149) is 5 times higher than that without common neighbors (0.00810). Figure 1(c) illustrates that the mobility similarity between non-friends is almost indifferently low no matter they have common neighbors or not. Befriending is strongly correlated with high mobility similarity between two individuals, with average $SCos$ increases by two orders of magnitude. Having common friends further doubles the similarity. Therefore, both "befriending" and "having common friends" imply high mobility similarity between individuals. Friends indeed are much closer in behavior pattern than strangers, and the existence of common neighbors could be another strong predictor of the similarity of individual mobility patterns. Because friends have a higher possibility of living or working together, or having the same hobbies, promoting the similarity of their mobility patterns compared with strangers. Meanwhile the common neighbors will strengthen the intimacy between friends. Besides, those two factors are affecting mobility similarity in different aspects and are not mutually replaceable.

In the above two metrics, common friends always provide richer information between two individuals compared with befriending, a rather intuitive criteria. For example, \emph{the number of common neighbors} ($CN$, or \emph{the size of common neighborhood}) are always regarded as an implication of intimate relationship in the research of link prediction\cite{lu2011} as local similarity indices and recommendation algorithm\cite{lu2012recommender} as structure similarity indices. Here, we count $CN$ of two individuals as the third measurement of social proximity (see Methods). One might expect that friends with more common neighbors have more similar mobility pattern, as suggested in previous studies\cite{Wang2011Human}. To our surprise, however, measurements from the following three aspects all reveal no positive impact of $CN$ on mobility similarity. To be specific, (1) A low \emph{Spearman coefficient}, 0.046, is observed between $SCos$ and $CN$, suggesting these two variables are less likely correlated numerically. (2) We compare the \emph{$SCos$ probability distributions} of 5 samples of friends (see Methods) with 1, 2, 3, 4, or 4+ common neighbor(s), namely $CN=1$, $CN=2$, $CN=3$, $CN=4$ or $CN\geq 4$. As shown in Fig. 1(d), friend pairs with different $CN$ (given $CN>0$) have almost identical distributions over mobility similarity, indicating that more common neighbors will not bring higher similarity. (3) \emph{Statistics hypothesis tests} are performed to statistically examine whether those samples are identically distributed. A null hypothesis that no difference lies between $CN=1$, $CN=2$, $CN=3$, $CN=4$ and $CN\geq 4$, is examined with $C_5^2=10$ pairwise $Kolmogorov-Smirnov$ statistics hypothesis tests. To exclude random error, each test is repeated for 1,000 parallel runs with Bonferroni correction (see Methods and Supporting Information), and none test rejects the null hypothesis. Therefore, we conclude that the mobility similarity is independent from the number of common neighbors since the samples are identically distributed. In a word, all the results consistently indicate the mutual independence between the mobility similarity and common neighbors. In another word, multiple common neighbors show equivalent effect as a single common neighbor when measuring mobility similarity. You have equal possibilities to find pairs with similar mobility patterns among those who are friends and having common friends, no matter how many common friends they have.

Shall we claim that common neighbor is a binary switch in shaping friends' mobility similarity while its details make no difference? No. It is the topological structure among common neighbors, instead of the size, that indicates greater mobility similarity. Define the common neighbor network of a pair of nodes as the induced subgraph by their common neighbors. Fig. 2(a) and (b) illustrate that there are various local organizations of common neighbor network when $CN$ is given. For example, with a certain $CN$, the common neighbors of a pair of individuals may cluster into a tightly connected group, or left isolated. Such phenomenon inspires us to investigate mobility similarity from the perspective of the micro-structure of common neighbors. Given the number of common neighbors, we measure the diversity of common neighborhood by \emph{the number of connected components} ($CC$). A higher $CC$ signifies more groups and higher degree of diversity of common neighbors. We collect individual pairs by configurations of $CN$ and $CC$, such as $\{CN=3, CC=2\}$. Figure 2(c) reports the average mobility similarity against common neighborhood size when we control its diversity, which, surprisingly, shows a consistently decreasing trend that more common neighbors lead to weaker mobility similarity. For example, if two individuals have 2 groups of 9 common neighbors, their mobility similarity could be as low as half of that when they have 2 groups of 2 common neighbors (i.e., two distinct common friends). On the other hand, as shown in Fig. 2(d), increasing diversity dramatically increases mobility similarity, when the number of common neighbors is controlled. Two individuals having 4 distinct friends (CC=4, CN=4) are twice similar in mobility than those having 4 connected friends (CC=1, CN=4). This phenomenon agrees with Ugander et al. \cite{ugander2012structural} that structural diversity of social network takes the role of common neighborhood size in shaping individual behaviors. It reveals that diversity of common neighbors is a signal of strong mobility similarity, while counter intuitively, the number of them give no positive effect.

\section*{Discussions}
~\\


Various kinds of human behaviors are highly correlated, from temporal to spatial, from online to offline. We analyzed the relation between online social proximity and offline mobility similarity in this work. Our empirical analysis reveals that mobility similarity between two individuals largely depends on their online social network connection, and further enhanced by the existence of common friends. Given the existence of common friends, the number of them shows no positive impact on mobility similarity. These results disagree with previous studies that believe the number of common friends is a positive predictor. It is worth noting that the number of connected components proves a consistent positive predictor of mobility similarity, though further experiments would be necessary to provide strong evidence.


The results can be explained from two aspects. On the one hand, it is not trivial to explain the phenomenon, namely individual mobility similarity is strongly related with the existence of common neighbors but hardly influenced by the number. Intuitively speaking, the phenomenon suggests that one common friend is enough to get a pair of individuals closer, while more common friends have no significantly additional effect. On the other hand, when the neighborhood of two individuals splits into pieces, there is a high probability that these two individuals belong to several different communities, strengthening their closeness simultaneously and leading to a higher similarity. From the perspective of human behavior analysis and link prediction, the existence of common neighbors is usually connected with direct friendship, but our experiments reveal that the existence of common neighbors and the direct friendship link affect offline mobility similarity respectively.



Different from previous studies that believe positive impact of the number of common neighbors, our empirically results show its negative impact while controlling the diversity in common neighbors. This could be explained in two folds. Firstly, the size and the number of connected components of common neighbors are trivially correlated, i.e., we cannot have 5 groups with less than 5 common friends. The auto correlation might lead to apparent positive relation between a larger number of common neighbors and a stronger mobility similarity, which indeed comes from the effect of diversity behind. Secondly, difference in data might also vary the conclusion. We collect data from individual check-in records, while previous studies leveraged mobile call GPS records\cite{Gonzalez2008Understanding, Cho2011Friendship}. The difference could arise from three facts. (1) Mobile call GPS records are coarser, reporting locations of base stations (separated by kilometers). Check-in records are finer, reporting coordinates of mobile device GPS (accuracy within $100$ meters). (2) Most human behaviors are trivial and less informative\cite{Song2010Limits}, resulting in noisy tracks reflected by purposelessly reported mobile call GPS data. In contrast, check-in records were submitted on purpose and thus believed to ensure a better signal-noise ratio. (3) Different kinds of social network reflects diverse social relationship. For example, Twitter is a directed network of follow-following relationship, while QQ in our research is an un-directed network with reciprocal relationships. There may be discrepancies in behaviors pattern aroused by the types of relationship, such as transferring information or causing behaviors.




Our analysis provides a statistical view of the coupling between human online social proximity and offline mobility similarity, and inspires deep understanding to the intrinsic of topological structure when predicting offline behaviors. Generally, the social network and check-in records correspond to real physical layer and virtual social layer in nature world and human society. Therefore, the LBSN data we used provides a good medium to couple physical space and social space. Technically, our results could offer new insight and evidence in the fields of location prediction and friend recommendation\cite{lian2015mining}. For example, mining human mobility pattern and leveraging social network information for next location prediction of a certain individual is always a big challenge. With deeper understanding of correlation between social proximity and mobility similarity, it will be easier to find someone ($A$)'s friend ($B$) who has more similar mobility pattern, which is helpful to predict $A$'s next location according to $B$'s trajectories. On the contrary, we could also recommend friend who has a higher mobility similarity with the target user to him/her.


Our study opens a door to a series of open questions. It is challenging and valuable to explain the fact that mobility similarity depends on the existence, but not the size, of online common neighbors. It remains unknown whether the effect of common neighbors could be generalized to more scenarios, before adequate empirical analysis is done on different types of social networks. It is also valuable to explore effect of common neighbors built with different types of edges, e.g., classmates, relatives, professional, etc.

\section*{Methods}

~\\
\textbf{{\em Data description}}

Our data is authorized by a Chinese online service provider Tencent, whose instant message product (QQ) and mobile check-in service provides the social network information and temporal-spatial mobility records respectively. QQ Users make friends and chat with them online as well as travel around offline in their daily life. Accordingly, on one hand, the social proximity between individuals can be depicted by the network structure. On the other hand, the trajectory sequence of each user and the similarity of mobility pattern between two individuals can be obtained as well. Therefore, this comprehensive dataset includes well coupled human online and offline behaviors. We were the first to analyze this dataset and will publish it with this paper.

Specifically, The users are sampled from a coastal city of China, while their check-in records cover the whole region of Chinese mainland. The dataset contains three parts of information, namely, individual demographic information, social relationship and time-stamped check-in records (longitude and latitude with an error no greater than $0.1km$). We remove inactive users with less than $100$ check-ins for stable statistics, resulting in a dataset of $97,657$ users with $617,765$ friend links and $28,827,898$ check-in records during the second half of year 2013. The average degree, average clustering coefficient and assortative coefficient of the social network is $6.32$, $0.09$ and $0.12$ respectively. As shown in Fig. \ref{fig:degree-distribution}, the degree distribution of social network follows power-law function with exponential cutoff.

~\\
\textbf{{\em Metrics definitions}}

In our research, $CN$ (number of common neighbors) is used as a metric of network proximity, which is defined as $CN(x,y)=|\Gamma(x)\cap\Gamma(y)|$, where $\Gamma(x)$ stands for the set of neighbors of individual $x$, $\Gamma(x)=\{y|y\in V,(x,y)\in E\}$, $E$ is the set of edges, $|A|$ is the size of set $A$. $SCos$ (spatial cosine similarity) is used as metric of mobility similarity by calculating the cosine similarity of two users¡¯ trajectory vectors as $SCos(x,y)=\sum_{l\in Loc}\frac{PV(x,l)¡¤PV(y,l)}{||PV(x,l)||\times ||PV(y,l)||}$, where $PV(x,l)$ stands for the probability of individual $x$ to visit location $l$, $||A||$ is the modulus of vector $A$. Only locations and its visit frequency is considered in this measurements, without considering the visiting sequence.

These two metrics measure the strength and similarity of social ties from online and offline aspects. We calculate $CN$ and $SCos$ for all pairs of friends in our dataset, and obtain the distributions of these two metrics which are shown in Fig.\ref{fig:metric-distributions}. It can been observed that both distributions derive from normal or approximate normal distribution. The results provide not only a description of heterogeneity of human behavior and society but also a evidence of choosing a appropriate correlation coefficient.

~\\
\textbf{\em{Sampling method and unbiased test}}

Due to the sparsity of social network (the network density in this study is only 0.00013), there are much more pairs of non-friends than friends. To reduce statistical error and computational complexity, we randomly select equal-sized pairs of non-friends as friends for the comparative study in the discrepancy of mobility similarity between them (Fig. 1(a) and (c)).

Considering that the degree distribution of our social network behaves heavy-tail shape, i.e., a few individuals have much more friends than majority, we sample individual pairs without overlapping to avoid auto-coupling. Specifically, we adopt the sampling without replacement method to pick out social ties to ensure that every individual appear in the sample only once to avoid the influence from the hub nodes with plenty of links. In the sampling process, once a edge is chosen, the two nodes connected by it will be removed from the sample pool. For example, individual k has friend i and j, if edge $(k,i)$ is chosen, both node $k$ and $i$ will be removed from nodes set and edge $(k,j)$ can't be used anymore. The obtained samples are used for investigating the mobility similarity of friends with different number of common neighbors.

The sampling is processed as follows. There are nearly 10 thousands nodes and over 600 thousands edges in the social network initially. Every step we pick out one edge $(i,j)$ randomly and the nodes $i$ and $j$ are removed from network. Finally some isolated nodes without any friend may be left and they are neglected. Thereby an ego-social network \cite{saramaki2014persistence, lu2013linked} is obtained where every edge represent a pair of individuals whose similarity is measured by $CN$ and $SCos$. The sampling process is carried out for 1000 times to ensure the randomness. In every sample, we have about 70 thousands nodes and nearly 40 thousands edges. Take 1 out of 1000 experiments as an example, 38,553 edges are obtained, within which 24,996 pairs of friends have no common neighbors and the remaining have at least one. In the same sample, the amounts of pairs whose $CN=$1,2,3,4 and $\geq$ 4 are 6459, 2903, 1555, 917 and 2640 respectively.

After sampling, unbiased test is used to ensure that all the samples keep consistent with each other. Specifically, we calculate the $mean$ and $std$ of $SCos$ and plot their distributions, which show a narrowed unimodal shape, illustrating that the 1000 samples are unbiased.

~\\
\textbf{{\em Kolmogorov-Smirnov test}}

\emph{Kolmogorov-Smirnov test} (KS test) is a kind of nonparameter test which can be used to verify whether two empirical samples are drawn from the same distribution. The KS statistic quantifies a distance between the empirical distribution functions of two samples. The null distribution of this statistic is calculated under the null hypothesis that the samples are drawn from the same distribution. The procedure of KS test can be seen in Ref. \cite{clauset2009power} and the Methods in Ref. \cite{yan2013diversity}. Besides, the null hypothesis is set as that the distribution samples are independent identically distributed (i.i.d.) with significance level $\alpha = 0.01$, thus the alternative hypothesis is that the samples are not i.i.d. In our research, owing to that the test is performed for 1000 times, the significance level is revised as adjust $\alpha = 0.01/1000 = 0.00001$ according to Bonferroni correction\cite{bonferroni1936teoria}. Therefore, as long as one of all the $p-values$ is smaller than adjusted $\alpha$, the null hypothesis is rejected and it can be deduced that two samples are different from each other. Conversely if all the $p-values$ are greater than adjusted $\alpha$, we can't reject the null hypothesis and the two samples are supposed to be extracted from the same population. In Supporting Information, we give the results of KS test of the distributions with different $CN$. If the tested two distributions are not i.i.d., the $p-values$ will follow a normal distribution. However, Fig.\ref{fig:ks-test-cn} demonstrates that all the distributions are not normally distributed, indicating that the two distributions are drawn from the same population.

\section*{Acknowledgments}
~\\
We thank Hao Chen (Nankai University), Huawei Shen (Chinese Academy of Sciences) and Simon DeDeo (Santa Fe Institute) for valuable comments. This work was partially supported by the National Natural Science Foundation of China (NNSFC) under Grant Nos. 61473060, 61433014, 61603074, 61673085, and the Hong Kong Scholars Program (No. XJ2013019 and G-YZ4D).

\section*{Author contributions}
~\\
CF, YL, JH, ZR and TZ designed research, CF and YL performed research, CF, YL, JH, ZR and TZ analyzed the data, and CF, JH and TZ wrote the paper.

\section*{Additional information}
\subsection*{Supplementary information}
accompanies this paper at http://www.nature.com/scientificreports

\subsection*{Competing financial interests}
The authors declare no competing financial interests.


\bibliographystyle{naturemag}
\bibliography{tencentmob}

\clearpage

\begin{figure}
\centering
\includegraphics[width=\textwidth]{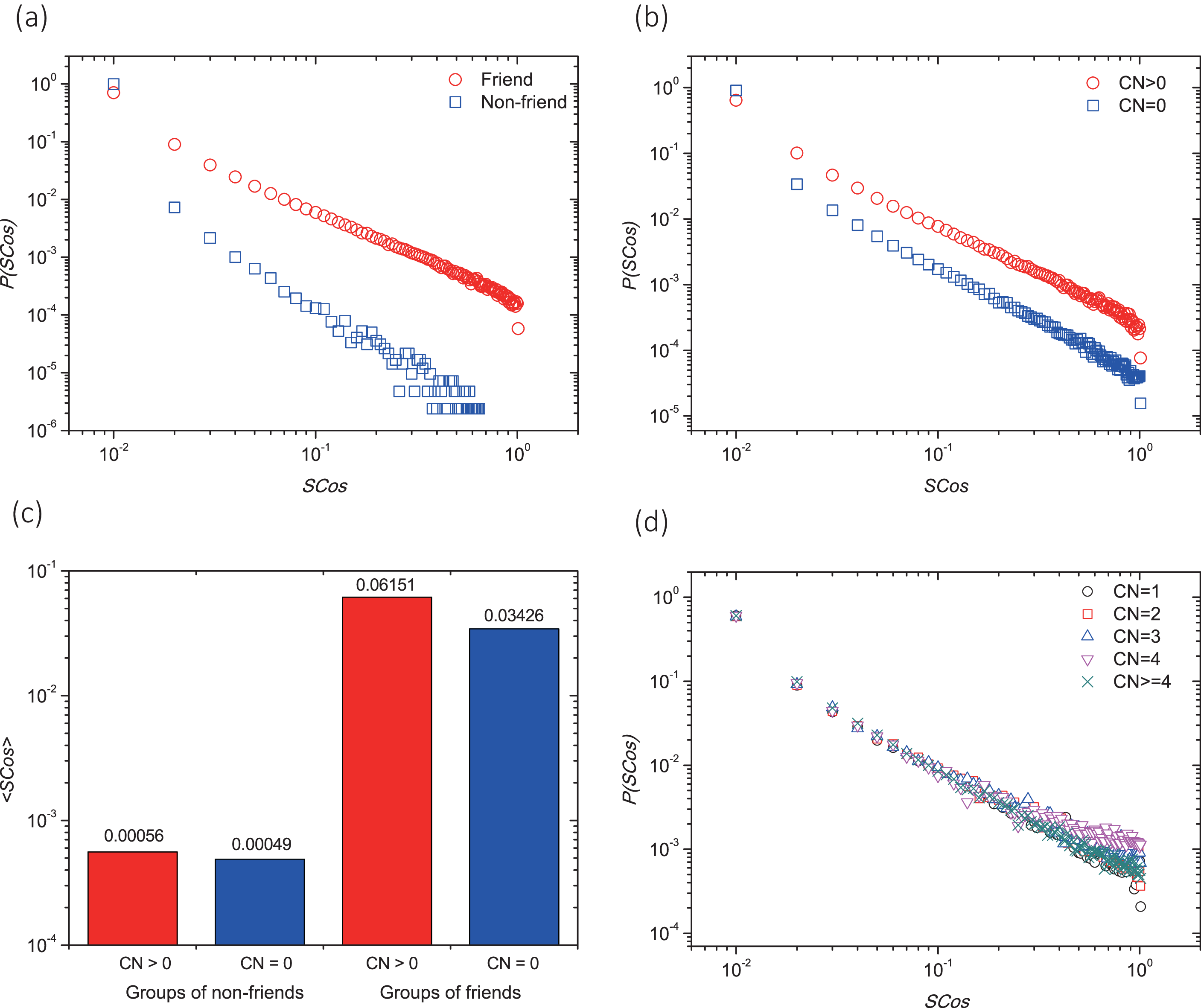}
\caption{\label{fig:effect-friends-commonfriends} \textbf{Correlation between befriending and having common friends with mobility similarity between two individuals.} (a), (b) and (d) show the probability distributions of Spatial Cosine Similarity (\emph{SCos}) for groups of different types of social relationship, namely, (a) pairs of individuals are or are not friends, (b) pairs of friends with or without common neighbors, and (d) pairs of friends binned by the different numbers of common neighbors. We add 0.01 to each data point to better illustrate zero in a log-log plot. In (a), it is consistently more probable to observe a pair of friends (red circles) with non-zero \emph{SCos} than a pair of non-friends (blue squares), while the former is much less probable to be observed with zero \emph{SCos}. Similarly in (b), the pairs with common neighbors (red circles) have higher mobility similarity than that without common neighbors (blue squares). However, almost invisible differences can be seen between the five groups of pairs with \emph{CN}=1,2,3,4 and $\geq$ 4 common neighbors in (d). In (c), the labels above the bars illustrate the average \emph{SCos} over all pairs of friends for 4 groups, by intersecting the two factors we observe. The differences between these 4 groups indicates that these two factors are not mutually inclusive. Notice that, we use logarithmic scale in (c) and thus the significant difference between red and blue bars are seemingly small.}
\end{figure}

\begin{figure}
\centering
\includegraphics[width=\textwidth]{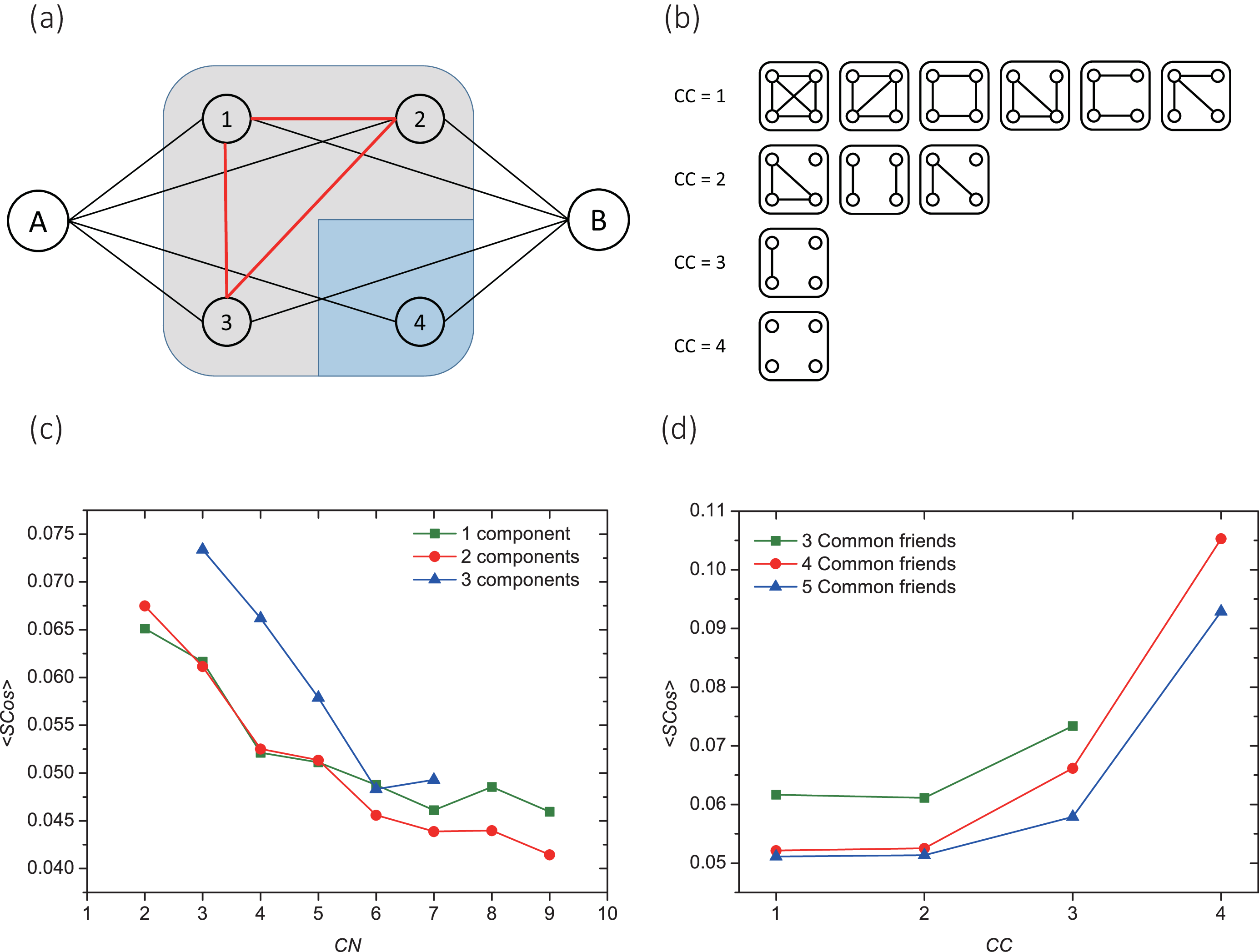}
\caption{\label{fig:effect-CN-CC-percolation} \textbf{Correlation between diversity of common friends with mobility similarity between two individuals.} (a) and (b) give an illustration of micro-structure of common neighbors. In (a), nodes 1, 2, 3 and 4 are common neighbors of nodes A and B. Nodes 1, 2 and 3 are wholly connected, while node 4 is isolated. Therefore, these 4 common neighbors are separated into 2 components. (b) shows all possible scenarios that 4 common neighbors may cluster into 1, 2, 3, or 4 connected components with different formation. (c) and (d) describe the average mobility similarity \emph{SCos} of samples in different configurations of the number of common neighbors (\emph{CN}) and connected components (\emph{CC}). Samples are grouped by the number of components (c) and common neighbors (d).}
\end{figure}

\clearpage


\setcounter{figure}{0}
\renewcommand{\thefigure}{S\arabic{figure}}

\begin{center}
\huge{Supporting Information}
\end{center}

\normalsize

\begin{figure}[!htb]
\centering
\includegraphics[width=0.8 \textwidth]{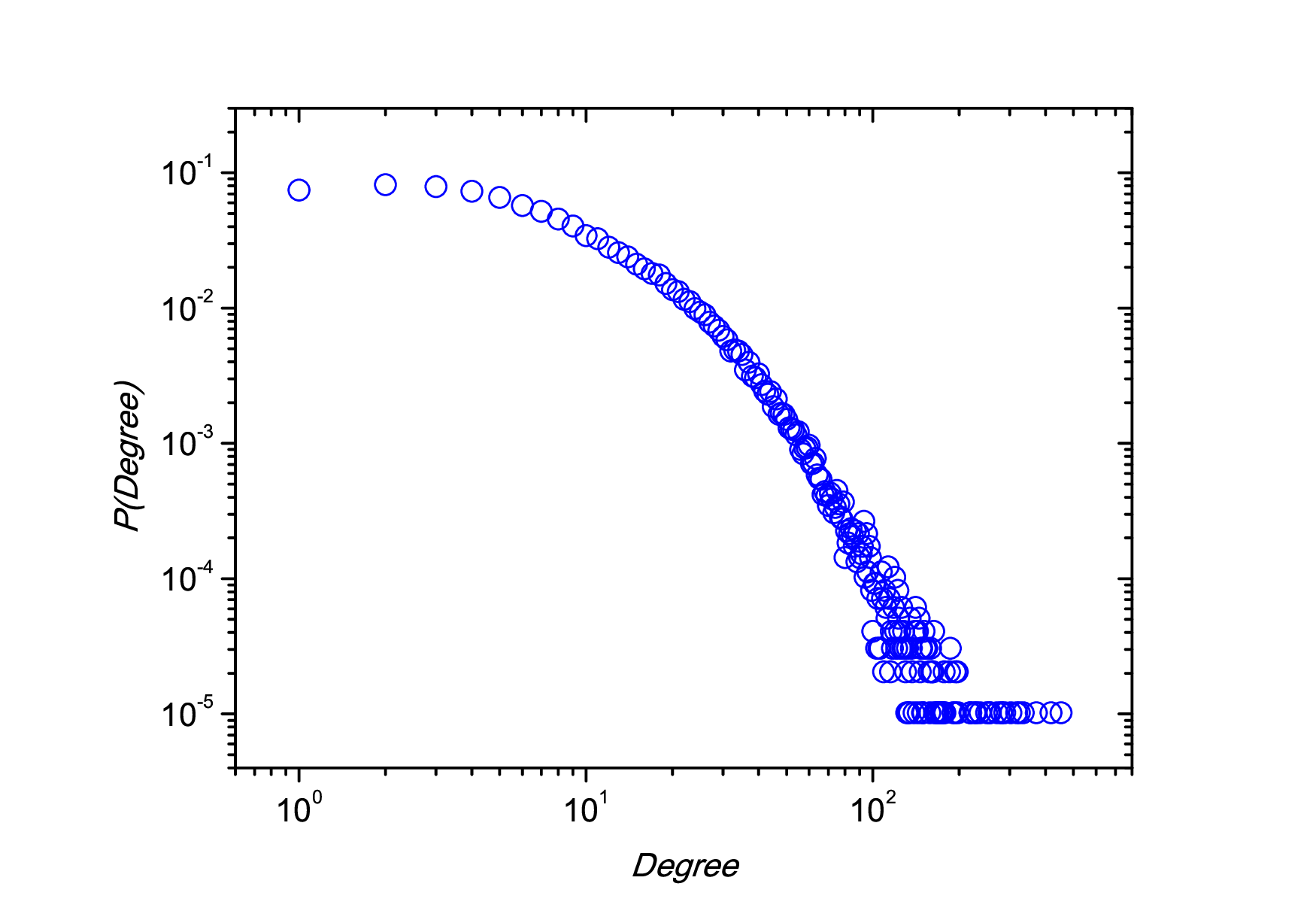}
\caption{\label{fig:degree-distribution} \textbf{The degree distribution of online social network.} It follows a heavy-tail shape which can be fitted by the function of power-law with exponential cutoff as: $f(x)\propto x^a * e^{-bx}$, where $a=0.157 \pm 0.027, b=0.121 \pm 0.005$..}
\end{figure}

\clearpage

\begin{figure}[!htb]
\centering
\includegraphics[width=0.8 \textwidth]{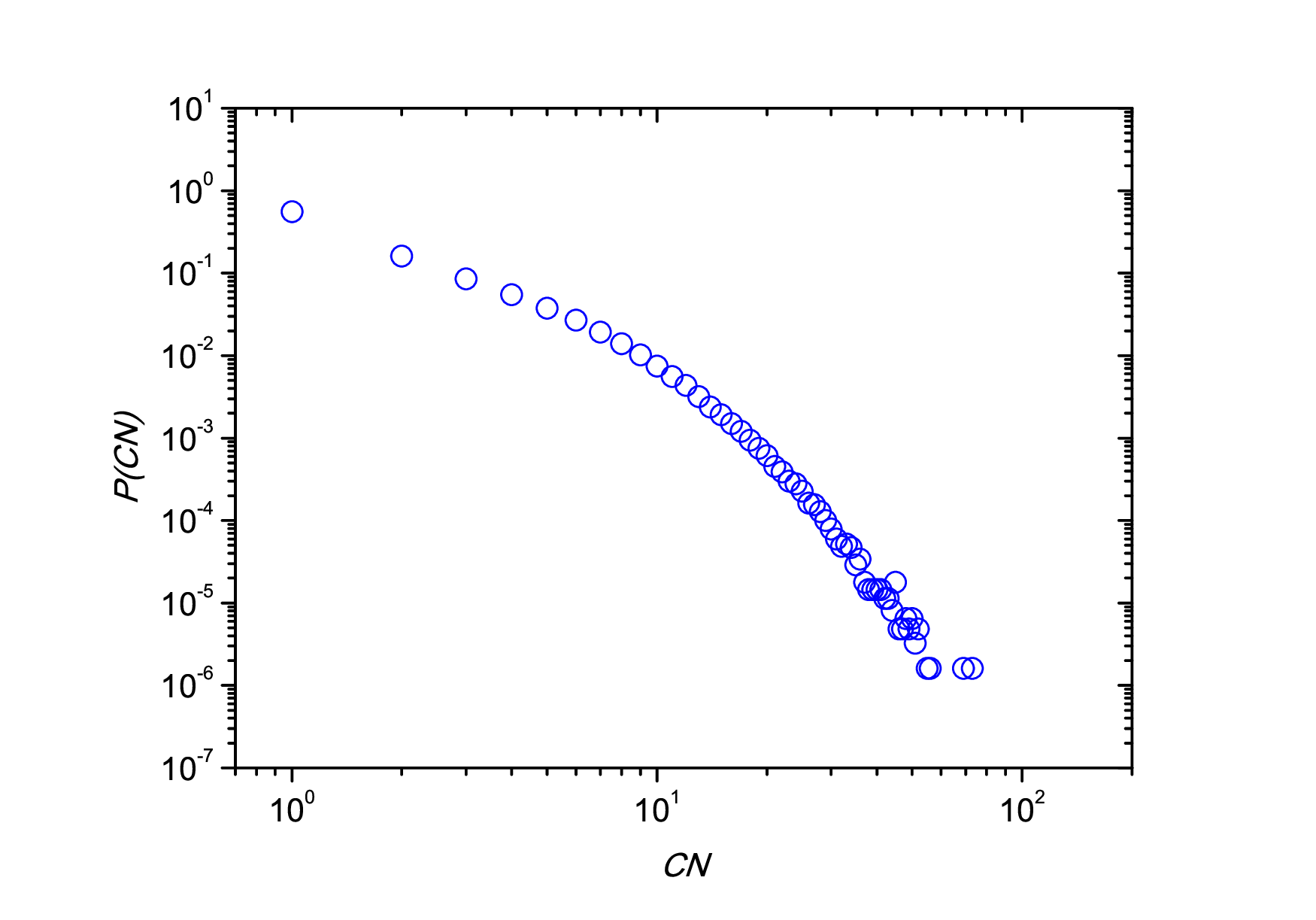}
\includegraphics[width=0.8 \textwidth]{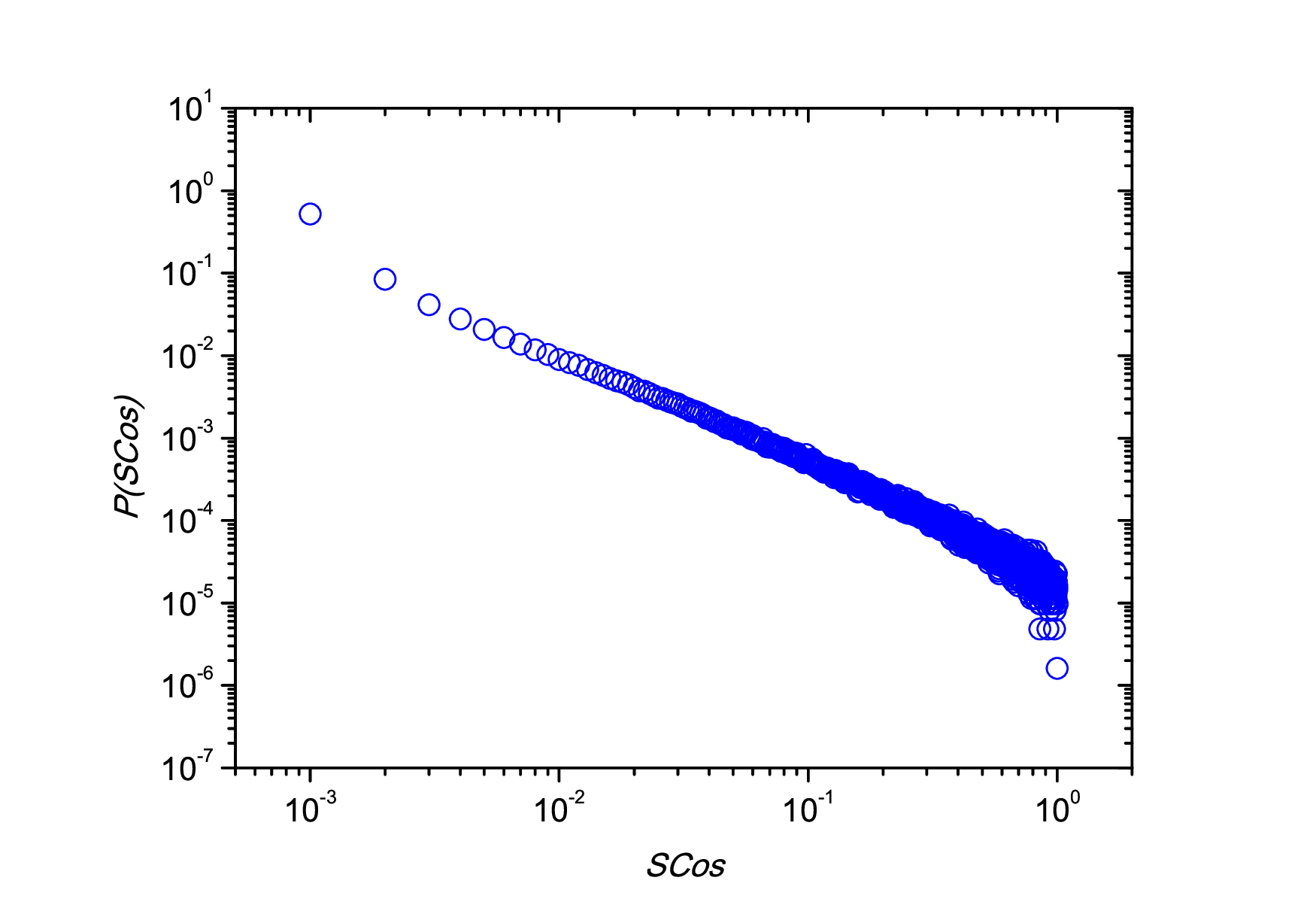}
\caption{\label{fig:metric-distributions} \textbf{The distributions of similarity metrics of \emph{CN} and \emph{SCos}.} Specifically, \emph{CN} follows the distribution of power-law with exponential cutoff as $f(x)\propto x^a * e^{-bx}$, where $a=-0.595 \pm 0.012, b=0.214 \pm 0.005$, and the distributions of \emph{SCos} behave power-law shape as $f(x)\propto x^a$, where $a=-1.041 \pm 0.003$. We add 0.001 for each data point of \emph{SCos} to better illustrate the zero values.}
\end{figure}

\clearpage

\begin{figure}[!htb]
\centering
\includegraphics[width=0.8 \textwidth]{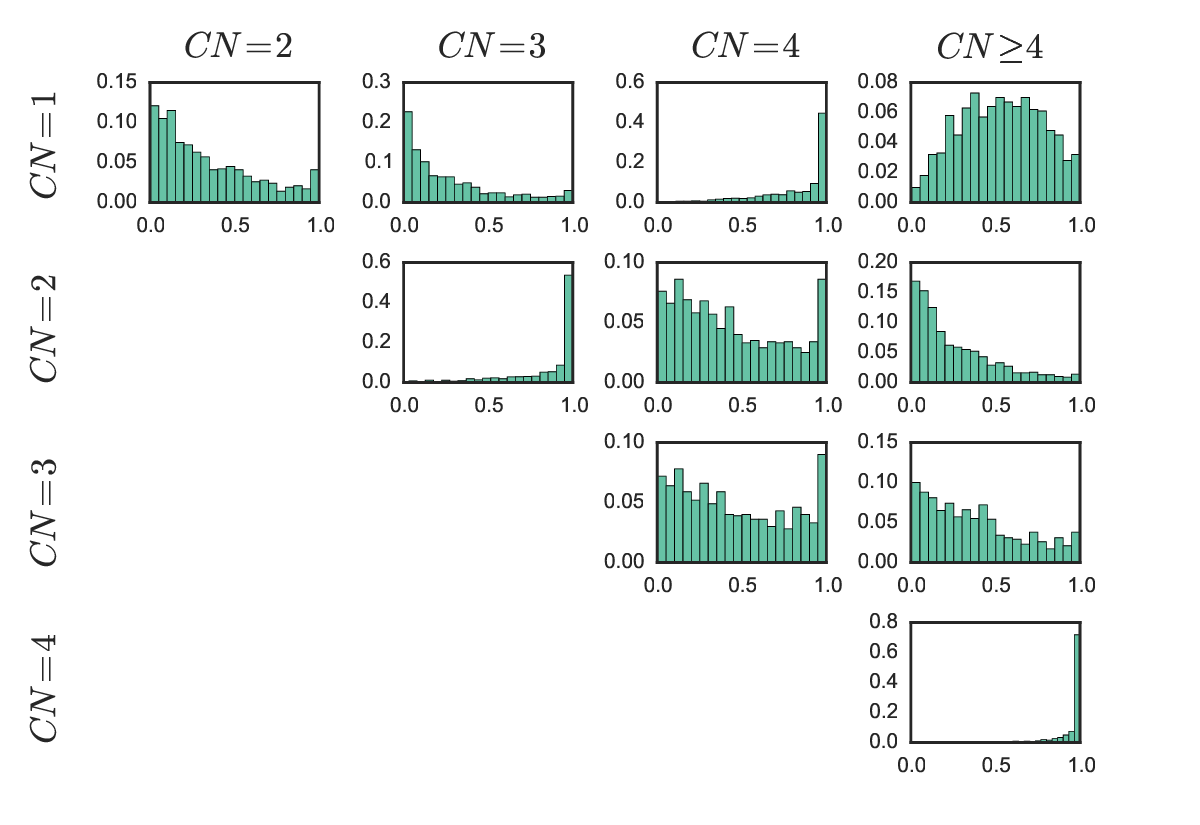}
\caption{\label{fig:ks-test-cn} \textbf{The distributions of \emph{p-value} of each KS test between groups with different \emph{CN}.} It's clear that all the distributions are not normally distributed, indicating that all pair of two distributions with different \emph{CN} are drawn from the same population.}
\end{figure}




\end{document}